\documentclass[pra,aps,amssymb,amsmath,amsmath,showpacs,reprint,twocolumn]{revtex4-1}
\usepackage{color}
\usepackage{graphicx}
\usepackage{epstopdf}
\usepackage{natbib}
\usepackage{hyperref}
\usepackage{dcolumn}
\usepackage{bm}
\usepackage{dcolumn}
\usepackage{multirow}

\newcolumntype{d}[1]{D{.}{.}{#1}}

\newcommand{\Fkt}[1]{\,\mathsf {#1}}
\def\openone{\leavevmode\hbox{\small1\kern-3.3pt\normalsize1}}

\ifx\Tr\renewcommand{\Tr}{\Fkt{Tr}} 
\else\newcommand{\Tr}{\Fkt{Tr}}
\fi

\usepackage{booktabs}
\usepackage{graphicx}
\usepackage{amsmath}
\usepackage{indentfirst}
\begin{document}

\title{Diamagnetic susceptibility of neon and argon including leading relativistic effects}

\author{\sc Micha\l\ Lesiuk}
\email{e-mail: m.lesiuk@uw.edu.pl}
\author{\sc Bogumi\l\ Jeziorski}
\affiliation{\sl Faculty of Chemistry, University of Warsaw\\
Pasteura 1, 02-093 Warsaw, Poland}
\date{\today}

\begin{abstract}
We report theoretical calculations of the static diamagnetic susceptibility, $\chi_0$, of neon and argon atoms. The calculations were performed using a hierarchy coupled-cluster methods combined with application of both the Gaussian and Slater orbital basis sets. We included the complete relativistic correction of order of $\alpha^4$, where $\alpha$ is the fine structure constant, and obtained an estimate of the quantum electrodynamics (QED) contributions. The finite nuclear mass and size corrections were also considered but are found to be small. The final results are $\chi_0=-8.4786(7)\cdot10^{-5}$ $a_0^3$ for the neon and $\chi_0=-22.9545(32)\cdot10^{-5}$ $a_0^3$ for the argon atom, where $a_0$ is the Bohr radius. The uncertainties in the last digits, shown in the parentheses,  
are primarily due to the errors in the non-relativistic electronic wavefunction, as well as due to the neglected quantum electrodynamics corrections.   
\end{abstract}

\maketitle

\section{Introduction}
\label{sec:intro}

In the preceding paper~\cite{part1}, we have reported accurate calculation of the static magnetic susceptibility, $\chi_0$, of helium atom taking into account the complete set of relativistic corrections as well as the finite-nuclear-mass effects. The magnetic susceptibility is a fundamental property of atoms and molecules that enables to determine their leading-order response (for closed-shell systems) to the applied external magnetic field. Moreover, this quantity is an important ingredient of the Lorentz-Lorenz formula~\cite{lorentz80,lorenz80}, which relates the refractive index, $n$, of an atomic gas with its density, $\rho$. The latter can be determined experimentally by measuring resonance frequencies of a quasi-spherical cavity under 
vacuum and when the cavity is filled with a working gas. Such measurements form the basis of the refractive-index gas thermometry (RIGT)~\cite{gao17,rourke19,ripa21,rourke21} -- a novel experimental technique in the field of metrology~\cite{jousten17,gaiser18,gaiser20,gaiser22}. Knowing the resonance frequencies in vacuum, $f(0)$, and at some pressure $p$ of the gas, $f(p)$, measured at a constant temperature, the refractive index is calculated as
\begin{align}
  n = \frac{f(0)}{f(p)\,(1-\kappa_{\mathrm{eff}}\,p)},
\end{align}
where $\kappa_{\mathrm{eff}}$ is an effective parameter, characteristic for a given apparatus, that accounts for the compression of the cavity as the pressure is increased. It is worth pointing out that this parameter is not affected by the composition of the working gas which opens up a window for an alternative experimental setup for RIGT measurements, as discussed by Schmidt \emph{et al.}~\cite{schmidt07} In this variant, two independent measurements are performed with two different working gases, such as helium and argon. The results are then combined to eliminate the $\kappa_{\mathrm{eff}}$ parameter entirely, thereby removing the uncertainty related to the compressibility of the cavity, see Ref.~\cite{rourke19}.

In the current realizations of the RIGT experimental setup, helium is the preferred working gas. This is justified by 
the accuracy of the theoretical data available for this system, in particular the 
polarizability~\cite{pachucki00,lach04,puchalski16,puchalski20}, magnetic 
susceptibility~\cite{Bruch:2002,Bruch:2003,part1}, density virial coefficients~\cite{cencek12,czachor20}, and 
dielectric/refractivity virial coefficients~\cite{rizzo02,cencek11,song20,garberoglio21}. However, the main disadvantage of helium is its small polarizability, making the RIGT measurements sensitive to small perturbations caused by inaccurate frequency or resonator compressibility determinations, and gas purity. The latter problem is especially troublesome as the most common impurity -- water vapour, is by roughly two orders of magnitude more polarizable than helium. For this reason, it has been suggested (see Ref.~\cite{rourke19} and references therein) to use other elements, in particular neon or argon, as the working gas. They have similar macroscopic properties as helium, but their polarizabilities are roughly by a factor of two and eight, respectively, larger. This helps to reduce the sensitivity of RIGT measurements to impurities and further improve their accuracy.

Despite the aforementioned advantages of neon and argon as the working gas, the knowledge of fundamental properties of these atoms is still incomplete. For example, their polarizabilities have only 
recently been calculated from first principles with high accuracy~\cite{lesiuk20,hellmann22,lesiuk23}. The magnetic susceptibility, $\chi_0$, of neon and argon are currently known with estimated error of several percent which is not satisfactory from the experimental point of view. In this work we report theoretical determination of $\chi_0$ for neon and argon following the the theoretical framework introduced in the preceding paper~\cite{part1}. For brevity, we refer to this work as Paper~I further in the text. We compute the complete set of relativistic contributions to $\chi_0$ (of the order of $\alpha^4$ where $\alpha$ is the fine-structure constant) and consider several other corrections due to finite nuclear mass and size, and quantum electrodynamics (QED) effects.

Atomic units (a.u.) are used throughout the present work, i.e. $\hbar\,$=$\,m_e\,$=$\,e\,$=$\,1$, where $m_e$ and $e$ are the electron mass and charge, respectively. We adopt the following  values~\cite{codata18} of fundamental physical constants: fine-structure constant, $\alpha=1/137.035\,999$, Bohr radius, $a_0 = 0.529\,177\,210\,$\AA{}, Avogadro number, $N_A=6.022\,140\,76 \cdot10^{23}$. The conversion factor between cm$^3$/mol, frequently used in the literature for $\chi_0$, and the atomic units is $1\,$cm$^3$/mol$=$11.205\,872  a.u.

\section{Theory}
\label{sec:theory}

For closed-shell atoms the diamagnetic susceptibility $\chi_0$ is defined as the second derivative of the energy, $E$, with respect to the strength $B=|\mathbf{B}|$ of the uniform external magnetic field $\mathbf{B}$, in the limit of $B\rightarrow 0$,
\begin{align}
 \label{chi0}
 \chi_0 = - \frac{\partial^2 E}{\partial B^2}\bigg|_{B=0}.
\end{align}
In general, the magnetic susceptibility is dependent on the frequency of the oscillating magnetic field. However, for closed-shall atoms the frequency-dependent terms appear only in the order of $\alpha^5$ and higher~\cite{yerokhin11,yerokhin12} or are quadratic in the electron-to-nucleus mass ratio, see the discussion in Ref.~\cite{lesiuk20}. As a result, the frequency contribution to $\chi_0$ is expected to be tiny and in this work we consider only static magnetic fields.

For closed-shell singlet electronic states, the dominant contribution (of the order $\alpha^2$) to $\chi_0$ is given by the formula~\cite{bethe75}
  \begin{align}
\chi_0^{(0)} = -\frac{1}{6 } \,\, \alpha^2  \, \langle \,\sum_i r_i^2  \, \rangle, 
\end{align}
where the summation index $i$ runs over all electrons in the system, $r_i=|\mathbf{r}_i|$ are the electron-nuclear distances, $\mathbf{r}_i$ are the spatial coordinates of the $i$th electron, and finally
$\langle X\rangle$ is a shorthand notation for the expectation value of an arbitrary operator $X$ with the non-relativistic ground-state electronic wavefunction, $\Psi_0$.
 
The relativistic corrections to $\chi_0$ of the order $\alpha^4$ can be divided into three groups. The first group comes from relativistic corrections to the electronic Hamiltonian resulting from Foldy-Woythausen expansion of the magnetic-field-dependent Dirac equation in powers of $\alpha^2$~\cite{pachucki:2008}. In Paper~I we have identified three corrections of this type which give diamagnetic contribution to $\chi_0$ and do not vanish after spin integration in a closed-shell system. They are given by the formulas:
\begin{align}
\label{chi1}
 \delta\chi^{(1)}_0 &= \frac{\alpha^4}{12}\,\langle \sum_i l_i^2 \rangle, \\
 \label{chi2}
 \delta\chi^{(2)}_0 &= -\frac{\alpha^4}{12}\,\langle \sum_i r_i^2\,\nabla_i^2 \rangle, \\
 \label{chi3}
 \delta\chi^{(3)}_0 &= \frac{\alpha^4}{4}N_e,
\end{align}
where $l_i^2$ is the square of the total electronic angular momentum operator for the $i$th electron, while $N_e$ denotes the number of electrons in the atom. These equations are equivalent to the formulas presented previously for the helium atom, cf.~Eq.~(13)-(15), and changes in the prefactors result solely from the use of atomic units here.

The second group of corrections originates from the Breit contribution to the electron-electron interaction. There are two corrections in this group, namely
\begin{align}
\label{chi4}
 \delta\chi^{(4)}_0 &= \frac{\alpha^4}{6}\,\langle \sum_{i<j} \frac{\mathbf{r}_i\cdot\mathbf{r}_j}{r_{ij}} \rangle, \\
 \label{chi5}
 \delta\chi^{(5)}_0 &= \frac{\alpha^4}{12}\,\langle \sum_{i<j} \frac{\mathbf{r}_i\cdot\mathbf{r}_j}{r_{ij}} -\frac{\big( \mathbf{r}_i\cdot\mathbf{r}_{ij} \big)
 \big( \mathbf{r}_j\cdot\mathbf{r}_{ij} \big)}{r_{ij}^3} \rangle,
\end{align}
where $\mathbf{r}_{ij}=\mathbf{r}_{i}-\mathbf{r}_{j}$, and $r_{ij}$ is the electron-electron distance.

Finally, the third group of corrections takes into account the relativistic corrections to the electronic wavefunction. Let us recall the standard form of the Breit-Pauli Hamiltonian~\cite{bethe75}
\begin{align}
\label{bph}
 \hat{H}_{\mathrm{BP}} = \hat{P}_4 + \hat{D}_1 + \hat{D}_2 + \hat{B},
\end{align}
where the above operators are defined as
\begin{align}
 \label{p4}
 \hat{P}_4 = -\frac{\alpha^2}{8}\,\sum_i \nabla_i^4,
\end{align}
\begin{align}
 \label{d1}
 \hat{D}_1 = \alpha^2\frac{\pi Z}{2}\, \sum_i \delta(\textbf{r}_{ia}),
\end{align}
\begin{align}
 \label{d2}
 \hat{D}_2 = \alpha^2\pi \sum_{i>j}\delta(\textbf{r}_{ij}),
\end{align}
\begin{align}
 \label{bb}
 \hat{B} = \frac{\alpha^2}{2}\sum_{i>j} \left[\frac{\nabla_i\cdot\nabla_j}{r_{ij}}
+\frac{\textbf{r}_{ij}\cdot(\textbf{r}_{ij}\cdot\nabla_j)\nabla_i}{r_{ij}^3} \right],
\end{align}
and $Z$ is the nuclear charge. Following the usual convention, we refer to these operators as the mass-velocity (MV), 
one-electron Darwin (D1), two-electron Darwin (D2) and orbit-orbit (OO), respectively. Every operator 
$\hat{X}=\hat{P}_4,\,\hat{D}_1,\,\hat{D}_2,\,\hat{B}$ appearing in the Breit-Pauli Hamiltonian gives an additional 
correction to the magnetic susceptibility of the following general form
\begin{align}
\label{chix}
 \delta\chi^{\mathrm{X}}_0 = -\frac{\alpha^2}{3}\langle\Psi_0| \Big(\sum_i r_i^2\Big)\frac{Q}{\hat{H}-E_0} \hat{X}|\Psi_0\rangle,
\end{align}
where $\hat{H}$ is the non-relativistic electronic Hamiltonian, $E_0$ is the ground-state electronic energy, and $Q=1-|\Psi_0\rangle\langle\Psi_0|$ denotes projection onto the subspace orthogonal to $\Psi_0$.

It is worth pointing out that besides the relativistic contributions to $\chi_0$, there are some minor corrections 
that account for the effects beyond the clamped-nucleus Born-Oppenheimer approximation: the finite nuclear size (FNS) 
and finite nuclear mass (FNM) corrections. They are considered in subsequent sections.

\section{Computational details}
\label{sec:kompot}

Calculation of the corrections listed in the previous section is a non-trivial problem and in many cases there are no 
programs available that can perform such task. In these cases, we developed and implemented the necessary formalism 
specifically for the purposes of this project. In this section we provide details of our calculations and specify the 
level of theory used to determine each contribution.

In the present work, two types of basis sets were employed in the calculations: Gaussian-type orbitals~\cite{gill94} (GTO) and 
Slater-type orbitals~\cite{slater30,slater32} (STO). The choice of the basis set type used in calculation of specific quantities was dictated 
by limitations of the available computer programs and by the accuracy required in the final results. In general, the 
available GTO for neon and argon are larger than the corresponding STO. Indeed, GTO for neon up to the bewildering 
tredecuple-zeta ($13$Z) quality are available from the recent work of Hellmann~\cite{hellmann22}. For argon, GTO up to nonuple-zeta ($9$Z) 
were optimized by us for calculations of the polarizability~\cite{lesiuk23}. In comparison, STO only up to $7$Z quality were reported 
for neon and argon~\cite{lesiuk20}. Therefore, we use GTO for calculation of $\chi_0^{(0)}$ where the accuracy requirements are the most 
stringent. Three program packages were used in these calculations: \textsc{Dalton}~\cite{daltonpaper}, \textsc{CFour}~\cite{cfour} and \textsc{MRCC}~\cite{kallay20} as 
detailed in the subsequent section. Additionally, GTO were used for calculation of the $\delta\chi^{\mathrm{MV}}_0$ and 
$\delta\chi^{\mathrm{D1}}_0$ corrections -- here we exploit the fact that such calculations can be performed with 
\textsc{Dalton} package without any modifications. Finally, GTO were employed in determination of the 
$\delta\chi^{\mathrm{D2}}_0$ and $\delta\chi^{\mathrm{B}}_0$ contributions which required to write a dedicated in-house 
program. The necessary orbit-orbit and two-electron Darwin integrals within GTO were exported from 
the \textsc{Dalton} package. To the best of our knowledge, a general implementation of the orbit-orbit integrals within 
STO is not available publicly.

The calculation of the remaining corrections, that is $\delta\chi^{(1)}_0$, $\delta\chi^{(2)}_0$, $\delta\chi^{(4)}_0$, $\delta\chi^{(5)}_0$, was accomplished within the STO. This choice is motivated by the fact that these corrections are not large and hence do not have to be determined as accurately. At the same time, we found that calculation of matrix elements corresponding to the operators appearing in Eqs.~(\ref{chi4}) and (\ref{chi5}) is actually simpler within STO than GTO (for atomic systems) which negates the main technical advantage of GTO. In fact, calculation of two-electron integrals within STO with the following interaction operators
\begin{align}
 K_1(\mathbf{r}_1,\mathbf{r}_2) &= \frac{\mathbf{r}_1\cdot\mathbf{r}_2}{r_{12}} \\
 K_2(\mathbf{r}_1,\mathbf{r}_2) &= \frac{\big( \mathbf{r}_1\cdot\mathbf{r}_{12} \big)
 \big( \mathbf{r}_2\cdot\mathbf{r}_{12} \big)}{r_{12}^3}
\end{align}
is a straightforward generalization of the formalism from Refs.~\cite{lesiuk14a,lesiuk14b} provided that partial-wave expansions (PWE) of these operators are available. Therefore, we seek the following PWE
\begin{align}
 K_i(\mathbf{r}_1,\mathbf{r}_2) &= \sum_{l=0}^\infty f_l^{(i)}(r_<,r_>)\,P_l(\cos\gamma),
\end{align}
for $i=1,2$, where $r_<=\min(r_1,r_2)$, $r_>=\max(r_1,r_2)$, and $\gamma$ is the angle between vectors $\mathbf{r}_1$ and $\mathbf{r}_2$. To derive the necessary expressions we first recall PWE for the Coulomb potential and for the interelectronic distance:
\begin{align}
\begin{split}
    &\frac{1}{r_{12}} = \sum_{l=0}^\infty\frac{r_<^l}{r_>^{l+1}}\,P_l(\cos\gamma), \\
    &r_{12} = \sum_{l=0}^\infty \Bigg[ \frac{1}{2l+3}\frac{r_<^{l+2}}{r_>^{l+1}}
        -\frac{1}{2l-1}\frac{r_<^l}{r_>^{l-1}}\Bigg]\,P_l(\cos\gamma).
\end{split}
\end{align}
With these formulas at hand, PWE for the first operator is obtained straightforwardly if we additionally exploit the law of cosines to eliminate $\mathbf{r}_1\cdot\mathbf{r}_2$:
\begin{align}
\begin{split}
 &K_1(\mathbf{r}_1,\mathbf{r}_2) =
 \frac{r_1^2+r_2^2}{2r_{12}} - \frac{1}{2}r_{12} =\\
 &\sum_{l=0}^\infty \Bigg[ \frac{l+1}{2l+3}\frac{r_<^{l+2}}{r_>^{l+1}}
 +\frac{l}{2l-1}\frac{r_<^l}{r_>^{l-1}}\Bigg]\,P_l(\cos\gamma),
\end{split}
\end{align}
The manipulations are somewhat more involved for the second operator. First, we recall the PWE for $r_{12}^{-3}$ which can be obtained as a special case of a more general formalism introduced by Sack~\cite{sack64}
\begin{align}
 r_{12}^{-3} &= \frac{1}{r_>^2-r_<^2}\sum_{l=0}^\infty(2l+1)\,
 \frac{r_<^l}{r_>^{l+1}}\,P_l(\cos\gamma),
\end{align}
After some algebra necessary to eliminate the vector quantities we find:
\begin{align}
\begin{split}
 &K_2(\mathbf{r}_1,\mathbf{r}_2) = \frac{\big(r_1^2 -r_2^2\big)^2}{4r_{12}^3}-\frac{1}{4}r_{12} = \\
 &\sum_{l=0}^\infty\Bigg[\frac{l^2}{2l-1}\,\frac{r_<^l}{r_>^{l-1}}-\frac{(l+1)^2}{2l+3}\,\frac{r_<^{l+2}}{r_>^{l+1}}\Bigg]\,P_l(\cos\gamma)
\end{split}
\end{align}
By inserting the PWE for $K_i(\mathbf{r}_1,\mathbf{r}_2)$ into the corresponding two-electron integrals within STO, the 
infinite summation over $l$ truncates and integration over all angles can be expressed through 3-$j$ symbols. The 
remaining radial integrals are simple linear combinations of the integrals encountered for the standard $1/r_{12}$ and 
$r_{12}$ operators, and hence no additional classes of basic integrals need to be implemented. In all calculations 
involving STO, a locally modified version of the \textsc{Gamess} package developed in Ref.~\cite{lesiuk15} was employed.

Calculation of one-electron integrals required to evaluate $\delta\chi^{(1)}_0$ and $\delta\chi^{(2)}_0$ is equally 
simple within the GTO and STO, and hence the choice of the latter was made for consistency.

\section{Numerical results}
\label{sec:num}

\subsection{The leading $\chi_0^{(0)}$ contribution}
\label{sec:chi0}

\begin{figure*}[ht]
\begin{tabular}{ccc}
 \includegraphics[scale=0.45]{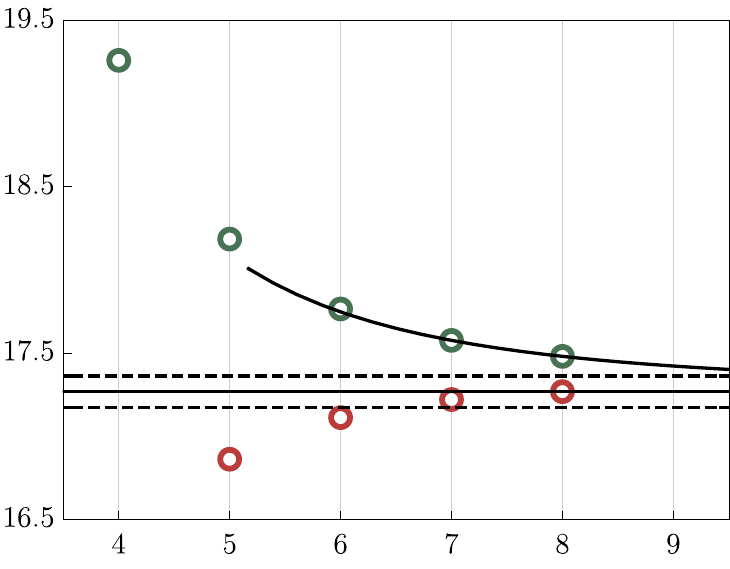} \hspace{0.3cm} & 
 \includegraphics[scale=0.45]{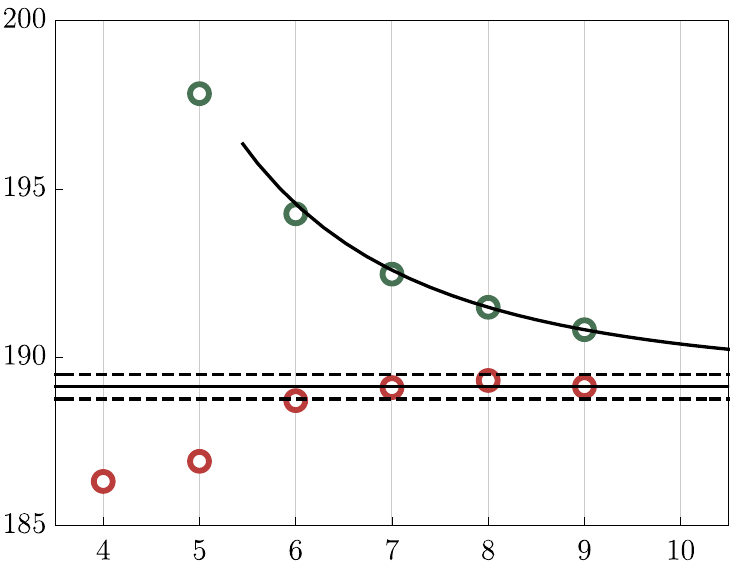} \hspace{0.3cm} &
 \includegraphics[scale=0.45]{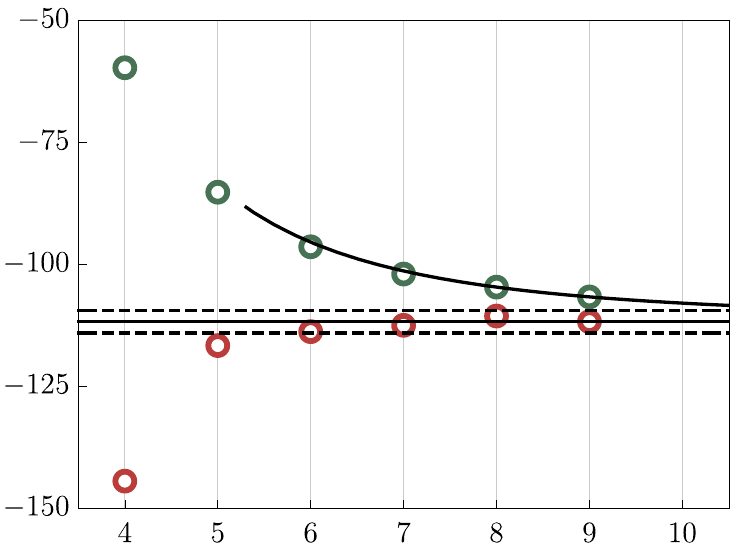} \\
\end{tabular}
\caption{\label{fig:cbs}
The corrections $\delta\langle r^2\rangle_{\mathrm{SD(T)}}$ for helium (left panel), neon (middle panel) and argon (right panel) atoms. Raw results obtained within dac$X$Z basis sets are represented by green points, while the values obtained by extrapolation from $(X,X-1)$ basis set pairs are given in red. The solid horizontal lines denote the best estimate obtained for $\delta\langle r^2\rangle_{\mathrm{SD(T)}}$ and the dashed horizontal lines represent the estimated error bars. For comparison, a curve $a+b/X^3$ obtained by fitting results from largest two basis sets is also plotted.
}
\end{figure*}

The $\chi_0^{(0)}$ contribution to the magnetic susceptibility is dominant and hence it has to be determined highly accurately. For this purpose we employ the doubly-augmented GTO, abbreviated shortly as d$X$Z further in the text, combined with a hierarchy of coupled-cluster (CC) methods~\cite{bartlett07,crawford07} which converge to the exact solution of the electronic Schr\"{o}dinger equation. The expectation value $\langle\sum_i r_i^2 \rangle$ entering $\chi_0^{(0)}$ is shortly denoted by the symbol $\langle r^2\rangle$ further in the text. This quantity is split into several components calculated at different levels of theory:
\begin{align}
\begin{split}
 \langle r^2\rangle &= \langle r^2\rangle_{\mathrm{HF}} 
 + \delta\langle r^2\rangle_{\mathrm{SD(T)}}
 + \delta\langle r^2\rangle_{\mathrm{T}} \\
 &+ \delta\langle r^2\rangle_{\mathrm{Q}}
 + \delta\langle r^2\rangle_{\mathrm{P}},
\end{split}
\end{align}
where the $\langle r^2\rangle_{\mathrm{HF}}$ is the Hartee-Fock contribution, while the remaining components $\delta\langle r^2\rangle_{\mathrm{X}}$ are corrections accounting for electron correlation effects obtained with the method $X$. For example, $\delta\langle r^2\rangle_{\mathrm{SD(T)}}$ is a correction to the Hartee-Fock result calculated using the CCSD(T) method~\cite{ragha89}, $\delta\langle r^2\rangle_{\mathrm{Q}}$ is the difference between CCSDTQ~\cite{kucharski91,kucharski92} and CCSDT~\cite{noga87,scuseria88} methods, an so on. The correction $\delta\langle r^2\rangle_{\mathrm{P}}$ was obtained using the CCSDTQP model~\cite{musial00,musial02}. Based on a set of preliminary calculations using CCSDTQPH~\cite{kallay01} and full configuration interaction (FCI) methods in small basis sets, we found that the contributions of CC excitations higher than pentuple are negligible. For all calculations reported in this subsection, we employed \textsc{CFour} program (CCSDT and lower-order methods) and \textsc{MRCC} program package (CCSDTQ and higher-order methods).

\begin{table}[b]
\caption{\label{tab:chi0}
The correction $\delta\langle r^2\rangle_{\mathrm{SD(T)}}$ obtained within the dac$X$Z basis sets family (all electrons correlated). In the case of helium, d2Z and d9Z basis sets are not available in the literature.
}
\begin{ruledtabular}
\begin{tabular}{cccc}
 \multirow{2}{*}{$X$} 
 & \multicolumn{3}{c}{$\delta\langle r^2\rangle_{\mathrm{SD(T)}}$} 
   \\[0.25em]\cline{2-4}\\[-1.1em]
     & He & Ne & Ar \\
 \hline\\[-1em]
 2 & ---     & 0.2549 & \phantom{$-$}0.1736 \\
 3 & 0.02253 & 0.2290 & \phantom{$-$}0.0330 \\
 4 & 0.01926 & 0.2067 & $-$0.0597 \\
 5 & 0.01819 & 0.1978 & $-$0.0852 \\
 6 & 0.01777 & 0.1943 & $-$0.0964 \\
 7 & 0.01758 & 0.1925 & $-$0.1020 \\
 8 & 0.01748 & 0.1915 & $-$0.1047 \\
 9 & ---     & 0.1908 & $-$0.1067 \\
 \hline\\[-1.2em]
 $\infty$ & 0.01727(9) & 0.1891(4) & $-$0.1117(23) \\[-0.1em]
\end{tabular}
\end{ruledtabular}
\end{table}

The Hartree-Fock contribution $\langle r^2\rangle_{\mathrm{HF}}$ is straightforward to calculate accurately using GTO, but even more accurate results are available in the literature from purely numerical HF computations based on $B$-splines expansion method. The following values were extracted from Ref.~\cite{saito09}:
\begin{align}
\label{r2hf}
\begin{split}
 &\mbox{He:}\;\;\langle r^2\rangle_{\mathrm{HF}}= 2.36\,966, \\
 &\mbox{Ne:}\;\;\langle r^2\rangle_{\mathrm{HF}}= 9.37\,184, \\
 &\mbox{Ar:}\;\;\langle r^2\rangle_{\mathrm{HF}}= 26.0\,344.
\end{split}
\end{align}
The values given above are accurate to all digits shown and hence the uncertainty of the $\langle 
r^2\rangle_{\mathrm{HF}}$ contribution is negligible. As a byproduct of subsequent calculations, we obtained $\langle 
r^2\rangle_{\mathrm{HF}}$ contributions within GTO for all atoms. Near-perfect agreement was obtained with the data 
given in Eq.~(\ref{r2hf}), differing only in the last digit in the case of neon and argon.

\begin{table}[b]
\caption{\label{tab:nehocc}
Corrections to $\langle r^2\rangle$ accounting for higher-order coupled-cluster excitations for neon atom (all electrons correlated). For clarity, the results were multiplied by a constant factor of $10^3$.
}
\begin{ruledtabular}
\begin{tabular}{cccc}
 $X$
 & $\delta\langle r^2\rangle_{\mathrm{T}}\cdot10^3$ 
 & $\delta\langle r^2\rangle_{\mathrm{Q}}\cdot10^3$ 
 & $\delta\langle r^2\rangle_{\mathrm{P}}\cdot10^3$ \\
 \hline\\[-1em]
2 & $-$0.268 & $-$0.333           & $-$0.328 \\
3 & $-$0.726 & \phantom{$-$}0.002 & $-$0.103 \\
4 & $-$1.068 & \phantom{$-$}0.104 & --- \\
5 & $-$1.219 & --- & --- \\
6 & $-$1.287 & --- & --- \\
 \hline\\[-1.2em]
 $\infty$ & $-$1.39(2) & \phantom{$-$}0.19(4) & \phantom{$-$}0.03(6) \\[-0.1em]
\end{tabular}
\end{ruledtabular}
\end{table}

The next large contribution to $\chi_0^{(0)}$ comes from the CCSD(T) level of theory, denoted $\delta\langle r^2\rangle_{\mathrm{SD(T)}}$. We do not adopt frozen-core approximation in these calculations and hence all electrons were correlated at this stage. In 
Table~\ref{tab:chi0} we report results of the calculations of the $\delta\langle r^2\rangle_{\mathrm{SD(T)}}$ 
contribution. To eliminate the remaining basis set incompleteness error, extrapolation towards the complete basis set 
(CBS) limit is required. To this end, we employ the formalism based on the Riemann zeta function~\cite{lesiuk19}. Let us 
assume that the quantity of interest $\mathcal{O}$ was calculated  with two consecutive basis sets ($X$ and $X-1$) and 
the results are denoted by symbols $\mathcal{O}_X$ and $\mathcal{O}_{X-1}$, respectively. The CBS limit 
$\mathcal{O}_\infty$ is then determined from the formula
\begin{align}
\label{riemann}
 \mathcal{O}_\infty = \mathcal{O}_X + 
 X^4\Big[\zeta(4)-\sum_{l=1}^X l^{-4}\Big]
 \big( \mathcal{O}_X - \mathcal{O}_{X-1} \big),
\end{align}
where $\zeta(s)=\sum_{n=1}^\infty n^{-s}$ is the Riemann zeta function and $\zeta(4)=\pi^4/90$. This 
extrapolation scheme shall be used for estimating CBS limits of all quantities considered in this work, unless 
extrapolation is deemed unnecessary. The determined CBS limits of the $\delta\langle r^2\rangle_{\mathrm{SD(T)}}$ 
contributions are given in Table~\ref{tab:chi0}. Clearly, the increments resulting from the extrapolation are sizeable 
and necessary to ascertain the reliability of the final data. To illustrate this, in Fig.~\ref{fig:cbs} we 
plot the calculated $\delta\langle r^2\rangle_{\mathrm{SD(T)}}$ corrections as a function of the parameter $X$. 
Additionally, we include the extrapolated results from basis set pairs $(X,X-1)$. The uncertainty of the extrapolation, 
represented in Fig.~\ref{fig:cbs} by horizontal dashed lines, is estimated as a difference between the CBS limits 
obtained with two largest basis set pairs (for example, $X=9,8$ and $X=8,7$ for neon and argon). One can see that the 
extrapolation is remarkably stable with respect to $X$. In particular, for neon and argon the last four extrapolated 
values already fall within the estimated error bars. This gives us confidence that the final value of $\delta\langle 
r^2\rangle_{\mathrm{SD(T)}}$ is not accidental and is supported by ample numerical evidence.

As an additional test, we analyze the data obtained for the helium atom and compare with the results reported in 
Paper~I. The latter are significantly more accurate and can be treated as a reference. By adding the $\langle 
r^2\rangle_{\mathrm{HF}}$ contribution from Eq.~(\ref{r2hf}) and the extrapolated $\delta\langle 
r^2\rangle_{\mathrm{SD(T)}}$ correction given in Table~\ref{tab:chi0}, we obtain $\langle r^2\rangle=2.38\,693(9)$. This 
compares favourably with the corresponding result from Paper~I, $\langle r^2\rangle=2.38\,697$, differing only at the 
last digit. Moreover, the difference is by a factor of two smaller than determined error bars, suggesting that our 
uncertainty estimation scheme is quite conservative. Moreover, it is worth pointing out that without extrapolation, i.e. 
by taking the $\delta\langle r^2\rangle_{\mathrm{SD(T)}}$ correction obtained within the largest basis set available in 
Table~\ref{tab:chi0}, one obtains $\langle r^2\rangle=2.38\,714$. The error of this quantity with respect to the 
reference data from Paper~I is more than four times larger than of the recommended extrapolated result. This shows that 
the adopted extrapolation scheme is reliable and enables to drastically reduce the residual basis set completeness 
error.

\begin{table}[t]
\caption{\label{tab:chi1}
Expectation values $\langle \sum_i l_i^2 \rangle$ obtained within the STO$X$Z basis sets family at the CCSD(T) level of 
theory (all electrons correlated).
}
\begin{ruledtabular}
\begin{tabular}{cccc}
 \multirow{2}{*}{$X$} 
 & \multicolumn{3}{c}{$\langle \sum_i l_i^2 \rangle$} 
   \\[0.25em]\cline{2-4}\\[-1.1em]
     & He & Ne & Ar \\
 \hline\\[-1em]
2 & 0.018\,375 & 12.0\,928 & 24.4\,445 \\
3 & 0.018\,740 & 12.1\,357 & 24.6\,352 \\
4 & 0.018\,860 & 12.1\,531 & 24.7\,157 \\
5 & 0.018\,913 & 12.1\,579 & 24.7\,329 \\
6 & 0.018\,934 & 12.1\,600 & 24.7\,393 \\
 \hline\\[-1.2em]
 $\infty$ & 0.018\,967(3) & 12.1\,632(7) & 24.7\,492(50) \\[-0.1em]
\end{tabular}
\end{ruledtabular}
\end{table}

\begin{table}[t]
\caption{\label{tab:chi2}
Expectation values $\langle \sum_i r_i^2\,\nabla_i^2 \rangle$ obtained within the STO$X$Z basis sets family at the 
CCSD(T) level of theory (all electrons correlated).
}
\begin{ruledtabular}
\begin{tabular}{cccc}
 \multirow{2}{*}{$X$} 
 & \multicolumn{3}{c}{$\langle \sum_i r_i^2\,\nabla_i^2 \rangle$} 
   \\[0.25em]\cline{2-4}\\[-1.1em]
     & He & Ne & Ar \\
 \hline\\[-1em]
2 & $-$0.140\,616 & 13.8\,928 & 38.1\,306 \\
3 & $-$0.139\,979 & 13.9\,613 & 38.4\,052 \\
4 & $-$0.139\,884 & 13.9\,867 & 38.5\,301 \\
5 & $-$0.139\,915 & 13.9\,943 & 38.5\,615 \\
6 & $-$0.139\,838 & 13.9\,980 & 38.5\,742 \\
 \hline\\[-1.2em]
 $\infty$ & $-$0.1397(2) & 14.0\,036(2) & 38.5\,939(63) \\[-0.1em]
\end{tabular}
\end{ruledtabular}
\end{table}

Next, we pass to determination of corrections to $\delta\langle r^2\rangle$ accounting for higher-order coupled-cluster excitations. In the case of argon, these corrections were determined in our recent paper~\cite{lesiuk23} and, despite significant effort, we did not manage to improve upon these results in a meaningful way. Therefore, we adopt the following values for argon:
\begin{align}
\begin{split}
 &\delta\langle r^2\rangle_{\mathrm{T}} = -0.0020(4), \\
 &\delta\langle r^2\rangle_{\mathrm{Q}} =\phantom{-}0.0032(12), \\
 &\delta\langle r^2\rangle_{\mathrm{P}} = -0.0012(2).
\end{split}
\end{align}
As one can see, these corrections accidentally nearly cancel out.

For neon, we carried out a new set of calculations of the corrections accounting for higher-order coupled-cluster excitations. The results are reported in Table~\ref{tab:nehocc}. In the case of the $\delta\langle r^2\rangle_{\mathrm{T}}$ contribution we managed to carry out calculations within Gaussian basis sets up to $X=6$. Therefore, we apply the same extrapolation and error estimation protocol as for the lower-order contributions. Unfortunately, the calculations of the $\delta\langle r^2\rangle_{\mathrm{Q}}$ and $\delta\langle r^2\rangle_{\mathrm{P}}$ corrections are feasible only within basis sets $X=2,3,4$ and $X=2,3$, respectively. Due to relatively small size of these basis sets, the results are not as reliable as the lower-level corrections. To account for this, the extrapolation is still performed using the formula~(\ref{riemann}), but the error is estimated as half of the difference between the extrapolated value and the result in the largest basis set available, see Table~\ref{tab:nehocc}.

\subsection{Relativistic corrections from the Dirac equation}
\label{sec:chi123}

In this section, we consider relativistic corrections to the magnetic susceptibility that originate from the expansion 
of the Dirac Hamiltonian, $\delta\chi^{(1)}_0$ and $\delta\chi^{(2)}_0$, defined in Eqs.~(\ref{chi1})-(\ref{chi2}). The 
remaining correction $\delta\chi^{(3)}_0$ is trivial to evaluate. Technically the simplest way of evaluating 
$\delta\chi^{(1)}_0$ and $\delta\chi^{(2)}_0$ is the finite-difference approach based on the Hellmann-Feynmann theorem
\begin{align}
\label{ffo}
 \langle \mathcal{O}\rangle = 
 \partial_\lambda\big|_{\lambda=0}\langle H + \lambda\mathcal{O}\rangle\approx
 \frac{\langle H + \lambda_0\mathcal{O}\rangle-\langle H - \lambda_0\mathcal{O}\rangle}{2\lambda_0},
\end{align}
where $\lambda_0$ is a suitably chosen small constant. The calculations were carried out at the all-electron CCSD(T) 
level of theory within STO denoted by the symbol STO$X$Z, where $X$ is the maximum angular momentum present in the 
basis. The STO were adopted from Ref.~\cite{lesiuk20} and include both the diffuse and core-valence functions.

In order to calculate the corrections $\delta\chi^{(i)}_0$, $i=1,\ldots,5$, using the finite-difference method, the 
operators in Eqs.~(\ref{chi1})-(\ref{chi5}) multiplied by the constant $\pm\lambda_0$ are added to the electronic 
Hamiltonian. The atomic energies evaluated with the modified Hamiltonians are used to extract the expectation values 
according to Eq.~(\ref{ffo}). This is straightforward in the case of all operators besides $\sum_i r_i^2\,\nabla_i^2$ 
which is not Hermitian and hence incompatible with the usual form of the electronic Hamiltonian. However, any operator 
can be written as a sum of a Hermitian and antihermitian operator as 
$\mathcal{O}=\frac{1}{2}\big(\mathcal{O}+\mathcal{O}^\dagger\big)+\frac{1}{2}\big(\mathcal{O}-\mathcal{O}^\dagger\big)$. 
Since expectation values of an antihermitian operator on electronic wavefunction vanish, it is sufficient to calculate only the expectation value of 
$\frac{1}{2}\big(\mathcal{O}+\mathcal{O}^\dagger\big)$ which is straightforward. Several values of the displacement 
$\lambda_0$ were tested in a preliminary calculations, but negligible difference were observed for $\lambda_0$ within 
the range $10^{-3}-10^{-5}$ and hence the midpoint of this interval, $\lambda_0=10^{-4}$, was used in all subsequent 
calculations. In general, the approximation~(\ref{ffo}) results in negligible errors in comparison to, e.g. basis set 
incompleteness.

\begin{table}[t]
\caption{\label{tab:chi4}
Expectation values present in the $\delta\chi^{(4)}_0$ correction, see Eq.~\ref{chi4}, obtained within the STO$X$Z basis sets family at the CCSD(T) level of theory (all electrons correlated).
}
\begin{ruledtabular}
\begin{tabular}{cccc}
 \multirow{2}{*}{$X$} 
 & \multicolumn{3}{c}{$\langle \sum_{i<j} r_{ij}^{-1}\,\mathbf{r}_i\cdot\mathbf{r}_j \rangle$} 
   \\[0.25em]\cline{2-4}\\[-1.1em]
     & He & Ne & Ar \\
 \hline\\[-1em]
2 & 0.06031 & 1.9286 & 6.7724 \\
3 & 0.05973 & 1.8514 & 6.4671 \\
4 & 0.05952 & 1.8237 & 6.3535 \\
5 & 0.05943 & 1.8158 & 6.3252 \\
6 & 0.05937 & 1.8123 & 6.3141 \\
 \hline\\[-1.2em]
 $\infty$ & 0.05929(3) & 1.8070(9) & 6.2968(64) \\[-0.1em]
\end{tabular}
\end{ruledtabular}
\end{table}

\begin{table}[t]
\caption{\label{tab:chi5}
Expectation values present in the $\delta\chi^{(5)}_0$ correction, see Eq.~\ref{chi5}, obtained within the STO$X$Z basis sets family at the CCSD(T) level of theory (all electrons correlated).
}
\begin{ruledtabular}
\begin{tabular}{cccc}
 \multirow{2}{*}{$X$} 
 & \multicolumn{3}{c}{$\langle \sum_{i<j} r_{ij}^{-1}\big(\mathbf{r}_i\cdot\mathbf{r}_j\big) -
 r_{ij}^{-3}\big( \mathbf{r}_i\cdot\mathbf{r}_{ij} \big)
 \big( \mathbf{r}_j\cdot\mathbf{r}_{ij} \big) \rangle$} 
   \\[0.25em]\cline{2-4}\\[-1.1em]
     & He & Ne & Ar \\
 \hline\\[-1em]
2 & 0.2129 & 7.4168 & 22.0487 \\
3 & 0.2128 & 7.3733 & 21.8951 \\
4 & 0.2127 & 7.3601 & 21.8494 \\
5 & 0.2126 & 7.3559 & 21.8374 \\
6 & 0.2126 & 7.3539 & 21.8325 \\
 \hline\\[-1.2em]
 $\infty$ & 0.2125(1) & 7.3509(2) & 21.8248(22) \\[-0.1em]
\end{tabular}
\end{ruledtabular}
\end{table}

In Tables~\ref{tab:chi1}~and~\ref{tab:chi2} we report results of the calculations of the expectation values $\langle 
\sum_i l_i^2 \rangle$ and $\langle \sum_i r_i^2\,\nabla_i^2 \rangle$, respectively. These operators are 
closely related to the kinetic energy operator and converge to the CBS limit at the same rate as the kinetic 
energy (and hence the total energy by the virtues of the virial theorem). Therefore, we extrapolated the results towards 
CBS using the formula~(\ref{riemann}) and the uncertainty was estimated in the same way as in Sec.~\ref{sec:chi0}.
Finally, we point out that the reference results obtained in our previous work for helium, $\langle \sum_i l_i^2 \rangle=0.018\,970\,526$ and $\langle \sum_i r_i^2\,\nabla_i^2 \rangle=-0.139\,689\,120$, agree with the values determined here, see Tables~\ref{tab:chi1}~and~\ref{tab:chi2}, within the estimated error bars of the latter.

The last correction, $\delta\chi^{(3)}_0$, depends only on the number of electrons in the system and is trivial to evaluate. We attach no uncertainty to this contribution.

\subsection{Relativistic corrections from the Breit interaction}
\label{sec:chi45}

Relativistic corrections originating from the Breit interaction, Eqs.~(\ref{chi4})~and~(\ref{chi5}), are somewhat more complicated than the contributions from the Dirac equation, because they involve two-electron operators. However, calculation of the corresponding matrix elements within the STO is manageable, as discussed in Sec.~\ref{sec:kompot}.
Fortunately, both operators present in Eqs.~(\ref{chi4})~and~(\ref{chi5}) are purely real and multiplicative, and hence Hermitian. Therefore, expectation values appearing in $\delta\chi^{(4)}_0$ and $\delta\chi^{(5)}_0$ can be calculated using the finite-difference approach, in an analogous way as in Sec.~\ref{sec:chi123}.

In Tables~\ref{tab:chi4}~and~\ref{tab:chi5} we report results of the calculations of the expectation values required in the $\delta\chi^{(4)}_0$ and $\delta\chi^{(5)}_0$ corrections, respectively. The computations were performed at the all-electron CCSD(T) level of theory. We adopt the same extrapolation and uncertainty estimation strategy as for the corrections originating from the Dirac equation, see Sec.~\ref{sec:chi123}. For helium, the results agree perfectly with the reference data from Paper~I.

\subsection{Relativistic corrections to the electronic wavefunction}
\label{sec:chid1}

\begin{table}[t]
\caption{\label{tab:chid1}
The correction $-\langle \big(\sum_i r_i^2\big)(H-E_0)^{-1}Q\,\hat{D}_1\rangle$ obtained within the da$X$Z basis sets family using the linear-response CC3 theory (all electrons correlated). The data was multiplied by a factor of $10^3$ for clarity.
}
\begin{ruledtabular}
\begin{tabular}{cccc}
 \multirow{2}{*}{$X$} 
 & \multicolumn{3}{c}{$-\langle \big(\sum_i r_i^2\big)(H-E_0)^{-1}Q\,\hat{D}_1\rangle\cdot 10^3$} 
   \\[0.25em]\cline{2-4}\\[-1.1em]
     & He & Ne & Ar \\
 \hline\\[-1em]
2 & ---       & 0.4888 & $-$0.3313 \\
3 & $-$0.4215 & 0.4856 & $-$0.4376 \\
4 & $-$0.4214 & 0.4560 & $-$0.7257 \\
5 & $-$0.4214 & 0.4475 & $-$0.8005 \\
6 & $-$0.4214 & 0.4436 & $-$0.8306 \\
7 & $-$0.4214 & 0.4416 & $-$0.8439 \\
8 & $-$0.4214 & 0.4404 & --- \\
 \hline\\[-1.2em]
 $\infty$ & $-$0.4214(1) & 0.4378(2) & $-$0.8690(83) \\[-0.1em]
\end{tabular}
\end{ruledtabular}
\end{table}

\begin{table}[t]
\caption{\label{tab:chip4}
The correction $-\langle \big(\sum_i r_i^2\big)(H-E_0)^{-1}Q\,\hat{P}_4\rangle$ obtained within the da$X$Z basis sets family using the linear-response CC3 theory (all electrons correlated). The data was multiplied by a factor of $10^3$ for clarity.
}
\begin{ruledtabular}
\begin{tabular}{cccc}
 \multirow{2}{*}{$X$} 
 & \multicolumn{3}{c}{$-\langle \big(\sum_i r_i^2\big)(H-E_0)^{-1}Q\,\hat{P}_4\rangle\cdot 10^3$} 
   \\[0.25em]\cline{2-4}\\[-1.1em]
     & He & Ne & Ar \\
 \hline\\[-1em]
2 & ---    & 2.786 & 29.46 \\
3 & 0.5304 & 2.800 & 29.34 \\
4 & 0.5318 & 2.852 & 29.65 \\
5 & 0.5324 & 2.872 & 29.74 \\
6 & 0.5327 & 2.880 & 29.78 \\
7 & 0.5329 & 2.885 & 29.80 \\
8 & 0.5331 & 2.888 & --- \\
 \hline\\[-1.2em]
 $\infty$ & 0.5334(1) & 2.893(5) & 29.83(2) \\[-0.1em]
\end{tabular}
\end{ruledtabular}
\end{table}

Relativistic corrections to the electronic wavefunction lead to additional contributions to the magnetic susceptibility, 
given by general formula~(\ref{chix}). Evaluation of these corrections requires some additional approximations. 
First, according to the results for helium from Paper~I, the contribution of the two-electron Darwin 
operator, $\delta\chi^{\mathrm{D_2}}_0$, is tiny. Note that this quantity involves the two-electron Dirac delta 
distribution which is sensitive only to the regions of the wavefunction where the electrons collide. This regime is 
governed by the Kato's cusp condition which is universal and does not depend on the system~\cite{kato57}. Therefore, we argue that 
the $\delta\chi^{\mathrm{D_2}}_0$ is small also for neon and argon, and neglect this correction from further 
considerations. It is worth pointing out that a similar phenomena was observed in calculations of the polarizability of 
noble gas atoms.

The corrections $\delta\chi^{\mathrm{D_1}}_0$ and $\delta\chi^{\mathrm{P_4}}_0$ are the dominant relativistic corrections of this type and need to be calculated accurately. For this purpose we employ the (orbital unrelaxed) linear response coupled-cluster theory based on the CC3 wavefunction as implemented in the \textsc{Dalton} program package. The corrections are obtained from the symmetric form of the polarization propagator at zero frequency as
\begin{align}
\begin{split}
 \delta\chi^{\mathrm{X}}_0 = \frac{\alpha^2}{6}\langle\langle X;r^2\rangle\rangle_{\omega=0}
\end{split}
\end{align}
where $r^2$ is a shorthand notation for $\sum_i r_i^2$, and $X$ is either the $D_1$ or the $P_4$ operator. The results obtained within the same GTO as used for in the $\delta\chi^{\mathrm{0}}_0$ calculations are given in Tables~\ref{tab:chid1}~and~\ref{tab:chip4}. The extrapolation towards the CBS limit and estimation of the uncertainty are performed according to the same protocol as for the previous contributions considered in Secs.~\ref{sec:chi123}~and~\ref{sec:chi45}.

Finally, we consider the orbit-orbit correction to the magnetic susceptibility, $\delta\chi^{\mathrm{B}}_0$. Taking into account the results for the helium atom reported in Paper~I, we expect this correction to be relatively minor. Therefore, it does not have to be computed as accurately as the contributions described in the previous paragraph. For simplicity, we adopt the Hartree-Fock approximation in calculation of $\delta\chi^{\mathrm{B}}_0$. The ground-state wavefunction is represented by a single Slater determinant, while the first-order response function, $(\hat{H}-E_0)^{-1}Q\big(\sum_i r_i^2\big)|\Psi_0\rangle,$ is expanded into a linear combination of singly-excited determinants. Contributions from higher-order excitations vanish due to Slater-Condon rules as $\sum_i r_i^2$ is a sum of one-electron operators.

The adoption of the Hartree-Fock wavefunction for $\Psi_0$ enables a significant truncation of the basis set used in the calculations. To expand the Hartree-Fock orbitals only basis set functions with angular momenta $l=0,1$ ($s$ and $p$) are needed. Additionally, since the operator $r_i^2$ is spherically symmetric, the same basis is sufficient also for the first-order response function. To saturate the results with respect to basis set size, we used a large GTO comprising $33s26p$ and $40s40p$ functions for neon and argon, respectively. The calculations were carried out using a program written specifically for this purpose and all necessary basic integrals were imported from a locally modified version of the \textsc{Dalton} package. We obtained
\begin{align}
 \begin{split}
 &\mbox{Ne:}\;\;\;\delta\chi^{\mathrm{B}}_0 = 0.00028(14), \\
 &\mbox{Ar:}\;\;\;\delta\chi^{\mathrm{B}}_0 = 0.0010(5),
 \end{split}
\end{align}
where we have adopted a large (50\%) uncertainty estimate to account for all approximations involved in the calculations.

\subsection{Estimation of higher-order QED contributions}

Besides the relativistic corrections to the magnetic susceptibility of the order $\alpha^4$, one has to consider higher-order corrections originating from quantum electrodynamics (QED). Rigorous calculation of these corrections is a formidable task beyond the scope of the present work. However, we can estimate the magnitude of the QED effects similarly as in Paper~I, namely by taking the relativistic correction which is the largest in magnitude and scaling it by the factor of $-\alpha\log(\alpha)\approx 0.036$. Both for neon and argon, the largest relativistic correction (in absolute terms) is $\delta\chi^{\mathrm{P4}}_0$. By scaling its value by $-\alpha\log(\alpha)$ we obtain the following estimates of the QED effects
\begin{align}
 \begin{split}
  &\mbox{Ne:}\;\;\;\delta\chi^{\mathrm{QED}}_0 = 0.0002(2), \\
  &\mbox{Ar:}\;\;\;\delta\chi^{\mathrm{QED}}_0 = 0.0019(19), \\
 \end{split}
\end{align}
where we have attached a very large ($100\%$) uncertainty to the resulting values.

\subsection{Estimation of the finite nuclear mass and size corrections}

Finally, let us discuss the finite nuclear size (FNS) and finite nuclear mass (FNM) corrections to the magnetic susceptibility. In order to estimate the former, the carried out calculations of the $\chi_0^{(0)}$ contribution using Gaussian finite nuclear model and compared with the same results obtained with point nucleus. Following the recommendations of Visscher and Dyall~\cite{visscher97}, we used a simple nuclear charge distribution in the form
\begin{align}
 \rho(r)=\rho_0\,e^{-\xi r^2}\;\;
 \rho_0=Z\Big(\frac{\xi}{\pi}\Big)^{3/2}\;\;
 \xi=\frac{3}{2\langle r_c^2\rangle},
\end{align}
where $\langle r_c^2\rangle$ is the averaged square of the nuclear charge radius which can be calculated for an isotope with atomic mass number $A$ from an empirical formula $\langle r_c^2\rangle=\big(0.836A^{1/9}+0.570\big)\,\mathrm{fm}$. The advantage of the Gaussian charge model is the fact that the corresponding electron-nucleus interaction potential is given by a simple expression
\begin{align}
 V(r)=-\frac{Z}{r}\mbox{erf}(\sqrt{\xi}r),
\end{align}
where $\mbox{erf}(x)$ is the error function. This form of the potential is trivial to incorporate into the standard quantum-chemical programs operating within Gaussian basis sets and the necessary integrals are available in the \textsc{LibInt} library~\cite{Libint2}. For simplicity, we applied the Gaussian nuclear model in the Hartree-Fock calculations of $\chi_0^{(0)}$. By comparing with analogous results obtained with point nucleus, we found that the FNS effects are of the order of $1\,$ppm for argon, and even less for neon. Therefore, they can be safely neglected in the present work.

The finite nuclear mass effects are also small for neon and argon. Even for $^4$He, these effects are not large, constituting about $4\cdot 10^{-4}$ of the total value of $\chi_0$, see Paper~I. For neon and argon, the FNM effects are expected to be smaller (on a relative basis) as they are dependent on the inverse of the nuclear mass. To estimate the magnitude of these corrections, we consider the dominant correction, i.e. the reduced-mass scaling term which is given by the formula
\begin{align}
\label{ms}
    \delta\chi_0^{\mathrm{FNM}} \approx \frac{3}{m_N}\,\chi_0^{(0)},
\end{align}
where $m_N$ is the nuclear mass. For the most abundant isotopes i.e.~$^{20}$Ne and $^{40}$Ar,
the mass-scaling corrections amounts to roughly $-7\cdot 10^{-4}$ and $-9\cdot 10^{-4}$, respectively.
Therefore, FNM effects become important for Ne and Ar only if relative accuracy levels better than $1\cdot10^{-4}$ are desired. Nonetheless, we include the $\delta\chi_0^{\mathrm{FNM}}$ correction in our final results and assign a large 50\% uncertainty to the corrections evaluated using Eq.~(\ref{ms}) in order to account for the missing terms, see Paper~I.

The small magnitude of the FNM and FNS corrections is straightforward to explain by noting that the dominant contribution to the magnetic susceptibility, namely $\chi_0^{(0)}$, involves the operator $\sum_i r_i^2$. This operator vanishes at the nuclear site and hence is relatively insensitive to minor changes of the electronic wavefunction in the vicinity of the nucleus resulting from FNM and FNS corrections.

\section{Summary and discussion}
\label{sec:concl}

\begin{table}[t]
\caption{\label{tab:summary}
Final results of the calculations of the static magnetic susceptibility, $\chi_0$, of neon and argon. For clarity, all terms are multiplied by a constant factor of $10^5$.
}
\begin{ruledtabular}
\begin{tabular}{lcc}
contribution & Ne & Ar \\
 \hline\\[-1em]
 $\chi_0^{\mathrm{HF}}$          
 & $-$8.3177\phantom{(0)} & $-$23.1061\phantom{(00)} \\
 $\delta\chi_0^{\mathrm{SD(T)}}$ 
 & $-$0.1678(4) & \phantom{$-$}0.0991(20) \\
 $\delta\chi_0^{\mathrm{T}}$     
 & \phantom{$-$}0.0012(1) & \phantom{$-$}0.0018(4)\phantom{0} \\
 $\delta\chi_0^{\mathrm{Q}}$     
 & $-$0.0002(1) & $-$0.0028(11) \\
 $\delta\chi_0^{\mathrm{P}}$     
 & $-$0.0000(1) & \phantom{$-$}0.0011(2)\phantom{0} \\
 \hline\\[-1.2em]
 \multicolumn{3}{c}{contributions from Dirac equation} \\
 \hline\\[-1em]
 $\delta\chi_0^{(1)}$ 
 & \phantom{$-$}0.0003(1) & \phantom{$-$}0.0006(1)\phantom{0} \\
 $\delta\chi_0^{(2)}$ 
 & $-$0.0003(1) & $-$0.0009(1)\phantom{0} \\
 $\delta\chi_0^{(3)}$ 
 & \phantom{$-$}0.0007\phantom{(0)} & \phantom{$-$}0.0013\phantom{(00)} \\[0.2em]
 \hline\\[-1.2em]
 \multicolumn{3}{c}{contributions from Breit interaction} \\
 \hline\\[-1em]
 $\delta\chi_0^{(4)}$ 
 & \phantom{$-$}0.0001(1) & \phantom{$-$}0.0003(1)\phantom{0} \\
 $\delta\chi_0^{(5)}$ 
 & \phantom{$-$}0.0002(1) & \phantom{$-$}0.0005(1)\phantom{0} \\[0.2em]
 \hline\\[-1.2em]
 \multicolumn{3}{c}{relativistic corrections to the wavefunction} \\
 \hline\\[-1em]
 $\delta\chi^{\mathrm{D1}}_0$ 
 & \phantom{$-$}0.0008(1) & $-$0.0015(1)\phantom{0} \\
 $\delta\chi^{\mathrm{P4}}_0$ 
 & \phantom{$-$}0.0051(1) & \phantom{$-$}0.0529(1)\phantom{0} \\
 $\delta\chi^{\mathrm{D2}}_0$ 
 & \multicolumn{2}{c}{neglected} \\
 $\delta\chi^{\mathrm{B}}_0$  
 & $-$0.0005(2) & $-$0.0018(9)\phantom{0} \\[0.2em]
 \hline\\[-1.2em]
 \multicolumn{3}{c}{other corrections} \\
 \hline\\[-1em]
 $\delta\chi^{\mathrm{QED}}_0$ 
 & \phantom{$-$}0.0002(2) & \phantom{$-$}0.0019(19) \\
 $\delta\chi^{\mathrm{FNM}}_0$ 
 & $-$0.0007(3) & $-$0.0009(4)\phantom{0} \\
 $\delta\chi^{\mathrm{FMS}}_0$ & \multicolumn{2}{c}{neglected} \\
 \hline\\[-1.2em]
 \multicolumn{3}{c}{total} \\
 \hline
 $\chi_0$ & $-$8.4786(7) & $-$22.9545(32) \\
\end{tabular}
\end{ruledtabular}
\end{table}

\begin{table*}[t]
\caption{\label{tab:literature}
Comparison with other theoretical and experimental literature values of static magnetic susceptibility of neon and argon. The error estimation is not present in cases where it has not been provided by the original authors. All values are given in the atomic units.}
\begin{ruledtabular}
\begin{tabular}{lll}
\multirow{2}{*}{literature reference} & \multicolumn{2}{c}{$\chi_0$}\\ \cline{2-3}
& \multicolumn{1}{c}{neon} & \multicolumn{1}{c}{argon} \\
\hline\\[-3.0ex]
\multicolumn{3}{c}{experimental} \\
\hline\\[-3.0ex]
 Havens~\cite{havens33}
 & $-8.574(9)\cdot 10^{-5}$ & $-2.15(2)\cdot 10^{-4}$ \\
 Mann~\cite{mann36} 
 & $-7.56(20)\cdot 10^{-5}$ & $-2.19(2)\cdot 10^{-4}$ \\
 \multirow{2}{*}{Barter \emph{el al.}~\cite{barter60}} 
 & \multirow{2}{*}{$-7.80(16)\cdot 10^{-4}$} & $-2.16(2)^a\cdot 10^{-4}$ \\
 &  & $-2.16(15)^b\cdot 10^{-5}$ \\
\hline\\[-3.0ex]
\multicolumn{3}{c}{theoretical} \\
\hline\\[-3.0ex]
 Yoshizawa and Hada~\cite{yoshizawa09}
 & $-8.27\cdot 10^{-5}$ & $-2.22\cdot 10^{-4}$ \\
 Ruud \emph{et al.}~\cite{ruud94} / Jaszu\'{n}ski \emph{et al.}~\cite{jaszunski95} 
 & $-8.48\cdot 10^{-5}$ & $-2.31\cdot 10^{-4}$ \\
 Reinsch and Meyer~\cite{reinsch76} 
 & $-8.57\cdot 10^{-5}$ & $-2.32\cdot 10^{-4}$ \\
 Levy and Perdew~\cite{levy85} / Desclaux~\cite{desclaux73} 
 & $-8.31\cdot 10^{-5}$ & $-2.30\cdot 10^{-4}$ \\
 Lesiuk and Jeziorski~\cite{lesiuk20,lesiuk23} &
 $-8.484(19)\cdot 10^{-5}$ & $-2.30(2)\cdot10^{-4}$ \\
\hline\\[-3.0ex]
this work & $-8.4786(7)\cdot10^{-5}$ & $-2.29548(32)\cdot10^{-4}$ \\
\end{tabular}
\end{ruledtabular}
\vspace{-0.3cm}
\begin{flushleft}
 $^a\,$original error estimate from Ref.~\cite{barter60}; \\
 $^b\,$revised error estimate proposed in Ref.~\cite{rourke21};
\end{flushleft}
\end{table*}

In Table~\ref{tab:summary} we gather all contributions to the magnetic susceptibility of neon and argon considered in this work, together with their respective uncertainties. The final values of $\chi_0$ are obtained as a sum of these contributions; the overall error is calculated by adding squares of the uncertainties in the individual components and taking the square root. This approach is justified by the standard error propagation formulas under the assumption that the uncertainties are not correlated in the statistical sense.

The final results determined in this work are $\chi_0=-8.4786(7)\cdot10^{-5}$ and $\chi_0=-22.9545(32)\cdot10^{-5}$ for neon and argon atoms, respectively. The relative uncertainty of the both quantities of the order of one part per ten thousand. In the case of neon, the total uncertainty is dominated by the $\delta\chi_0^{\mathrm{SD(T)}}$ contribution, amounting to more than 50\% of the overall error. It is possible that the accuracy of this component can be improved in the future by using even larger Gaussian basis sets in the calculations and/or adopting a different extrapolation scheme. However, to reduce the total uncertainty by an order of magnitude, two other contributions, namely $\delta\chi^{\mathrm{B}}_0$ and $\delta\chi^{\mathrm{QED}}_0$, have to be determined more accurately. In the case of $\delta\chi^{\mathrm{B}}_0$ this would require a treatment with inclusion of electron correlation. Fortunately, the correlation contribution to $\delta\chi^{\mathrm{B}}_0$ appears to be small, so low-level methods such as MP2~\cite{moller34} or CC2~\cite{christiansen95} may be entirely sufficient, avoiding technical complications of higher-order methods. To improve the relative accuracy by an order of magnitude, finite nuclear mass effects must also be determined for neon. Finite nuclear size contributions are negligible up to $1\,$ppm uncertainty level.

The dominant sources of uncertainty are similar for argon, but are somewhat larger in magnitude. In particular, due to larger nuclear charge, the errors resulting from $\delta\chi^{\mathrm{B}}_0$, $\delta\chi^{\mathrm{QED}}_0$, and $\delta\chi_0^{\mathrm{SD(T)}}$ are comparable in magnitude. Therefore, further improvements in accuracy would require a more rigorous treatment of the $\delta\chi^{\mathrm{QED}}_0$ contribution. Both for neon and argon, there is an additional source of uncertainty coming from the $\delta\chi_0^{\mathrm{T}}$, $\delta\chi_0^{\mathrm{Q}}$, and $\delta\chi_0^{\mathrm{P}}$ corrections. However, these contributions cancel to a significant degree, so their impact on the overall accuracy is not expected to be large.

This analysis leads to the conclusion that it is worthwhile to derive and evaluate the complete $\delta\chi^{\mathrm{QED}}_0$ correction for all noble gas atoms, as it constitutes the main source of uncertainty that cannot be reduced by, e.g. using a larger basis set. While determination of the QED corrections to the energy is now possible~\cite{pachucki06,piszcz09,cencek12,lesiuk15,balcerzak17,lesiuk19a,lesiuk23,lesiuk23b}, this is not the case for atomic and molecular properties such as polarizability or magnetic susceptibility. In fact, the required theoretical framework has not been developed yet and this would require a considerable progress beyond the current state of the art.

In Table~\ref{tab:literature} we compare out results with the available theoretical and experimental data. The previous theoretical determination are in rough agreement with our data, but are significantly less accurate. However, the present results are in disagreement with the experimental data of Barter~\emph{et al.}~\cite{barter60}. In the case of argon, the value given by Barter~\emph{et al.}~\cite{barter60} is an average of three previous measurements (used to calibrate the apparatus) and does not count as an independent experimental determination. It has been suggested~\cite{rourke21} that the experimental uncertainty for argon has to be increased to about 7\%. After this revision, theory and experiment agree for argon, but for neon a large discrepancy remains. Somewhat unexpectedly, for neon a good agreement is obtained with older experimental data by Havens~\cite{havens33}. The reasons for the observed disagreement with the work of Barter~\emph{et al.}~\cite{barter60} are not known definitively. In Paper~I we discussed possible sources of the discrepancy in the case of helium, and similar conclusions apply to neon and argon. In short, physical effects neglected in our calculations are most likely orders of magnitude too small to explain such large discrepancy. We uphold our conviction that new and independent measurements of the magnetizability of the noble gases are needed to resolve this issue.

To sum up, we have reported state-of-the-art calculations of the static magnetic susceptibility of neon and argon. The results appear to be the most accurate available in the literature. They will be useful, for example, in refractive-index gas thermometry measurements or as a benchmark for other theoretical methods.

\begin{acknowledgments}
We thank M. Przybytek (UW) for providing several types of integrals required in the relativistic calculations. This project (QuantumPascal project 18SIB04) has received funding from the EMPIR programme cofinanced by the Participating States and from the European Union’s Horizon 2020 research
and innovation program. The authors also acknowledge support from the National Science Center, Poland, within the Project No. 2017/27/B/ST4/02739. We gratefully acknowledge Poland's high-performance Infrastructure PLGrid (HPC Centers: ACK Cyfronet AGH, PCSS, CI TASK, WCSS) for providing computer facilities and support within computational grant PLG/2023/016599.
\end{acknowledgments}

\bibliography{chi_ne-ar}

\end{document}